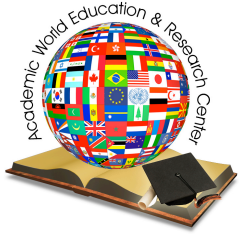

# The importance of human-computer interaction in the development process of software projects

**M. Hanefi Calp \***, Gazi University, Department of Management Information Systems, Institute of Informatics, Gazi University, Ankara, 06500, Turkey.

**M. Ali Akcayol,** Department of Computer Engineering, Faculty of Engineering, Gazi University, Ankara, 06500, Turkey.

## Abstract

Today, software industry has a rapid growth. In order to resist the competition increased by this growth, software projects ne0ed to be developed with higher quality and especially user friendly. Therefore, the importance of human-computer interaction emerges clearly. In design and development phases of software projects, the properties of human –which is an important agent for interaction- such as behavioral, cognitive, perceptive, efficiency and physical factors have to be considered. This study aims to express the importance of developing softwares by taking into consideration the human-computer interaction applications. In this context, firstly a wide literature review is made to examine software development process and human-computer interaction in detail, the results obtained by using design methods in this process are explicated and the importance of said interaction is openly expressed with the exemplary applications in the literature. According to the results of the research, especially in software life cycle, it is observed that rules of interaction must be implemented before software development, however, these methods are usually included in software life cycle in the latter stages of software development process. This situation causes the developed softwares to be user unfriendly and of low quality. Furthermore, it is observed that when the design methods used in the scope of human-computer interaction are integrated into software development process during the life cycle, the developed projects are more successful, have better quality and are more user friendly.

\*ADDRESS FOR CORRESPONDENCE: **M. Hanefi Calp,** Gazi University, Department of Management Information Systems, Institute of Informatics, Gazi University, Ankara, 06500, Turkey. *E-mail address*: **hcalp25@hotmail.com**

## 1. Introduction

Software industry has become very important with taking place in almost every area of human life and the size of the developed products. The software developed by coders are used in both production and consumption areas from our mobile phones to computers, from the citizenship process to the health sector, from the military to energy. With each passing day it becomes more widespread and indispensable, developing such software accurate and usable form is important for software companies [1, 2].

To achieve this, Software Engineering (SE) and Human-Computer Interaction (HCI) standards should be taken seriously implemented. In recent years, SE and HCI has made major improvements to meet customers' needs and responsibilities. These issues not only in themselves but also have provided important contribution to the understanding and development of each other. SE are used to get the requirements, to understand the needs of users and other stakeholders and, especially in the design phase, to develop process models. HCI is a discipline used to demonstrate technical capabilities and constraints of software designs of developers, and also developed the software prototypes for user reviews [3, 5].

Therefore, in order to create an effective and successful software product, sense of the demands and needs of users and the above-mentioned topics should be taken into consideration by developers. In this context, the usability is really important in terms of human-computer interaction. Usability is a concept originated from HCI that makes people learn a system developed easily, ensures that the system includes necessary functions that facilitates people's job and the system is easy to use and satisfying. At the same time, the user performance from the software development perspective is a software quality attributes corresponding satisfaction and learnability [6, 8]. The usability concept and software development is closely related in terms of the success of developed products. In the literature, there are many studies referring to integration of usability to software life cycle [6, 8, 12]. However, the usability concept are differently perceived by HCI and SE fields although it is a concept that is intersection of both areas. Software development is mainly associated with system functions but usability focus on the user mostly. The user is very important in both areas, but there are differences about the role of user. The user is seen as the primary source of requirements in the software development and the primary source of the system design in usability side [6, 13]. As a result, giving importance to the HCI in the Software Development Process (SDP) will provide an important contribution to –friendliness of the user, usability and more quality of the developed software.

In this study, Human-Computer Interaction in SDP are given in the second part (software development process and human-computer interaction) and finally the results and suggestions obtained from the study in the third part.

## 2. Human-Computer Interaction in Software Development Process

With the literature, the relationship SDP and HCI disciplines and the integration of these disciplines were investigated in manyresearch [8, 12, 14, 17]. Given the tendency for integration and that erros caught earlier in the development process are cheaper to fix compared to errors caught later in the process, in the same manner, it s thought that usability could be assessed before starting to develop a product and during design and requirement analysis [14]. At this point, the SDP and HCI issues is required to address separately in terms of better understanding of the importance of HCI in the development process of software projects.

## 2.1. Software Development Process

SDP is a set of actions associated with the development of a software product. The development process is basically a set of activities aimed to developed or evaluated of the software. The main goal in SDP is to prefer a good work in design rather than a good work in test, (to use the design tool instead of the debugger), to provide the quality preventing not correcting errors, to see the maintenance as error preventive, to update simultaneously the design and requirements, to prefer focus the process instead of focus the activity separately, to do not hesitate to build cross-functional (cross-functional) teams and to add customers this team, to distribute the responsibility to the entire of team [2, 18, 21]. General steps of the software development process is shown in Figure 1.

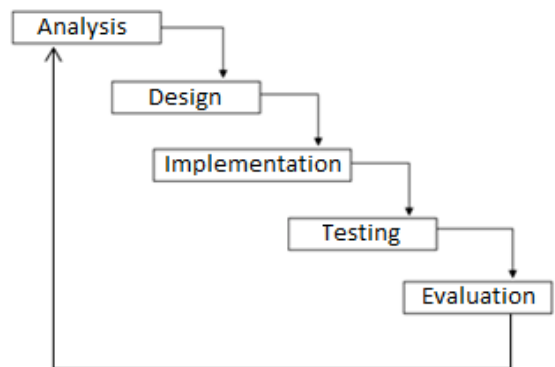

Figure 1. Software Development Process

- *Analysis* defines the resolution of the problem considered to be overcome. In this section; there are various requirement documents containing the requirements of a problem in complete, accurate, clear and understandable manner.
- *Design* describes detailed structures and properties of components of the system model. In this section, data structures and algorithms in each component of the model is described in detail.
- *Implementation* includes the change of the design solutions in programming language. This is the real structure of the program(s) in application. This section is a set of created programs.
- *Testing* confirms whether the programs are working, or not according to the obtained sample set and requirements with valid data.
- *Evaluation* includes the evaluation of results and, installation and delivery of program for users [18, 22].

## 2.2. Human-Computer Interaction

Human factors must be considered firstly in order to analyze of human-computer interaction issues well. Because human is the primary factor in HCI, interactive communication, cognitive skills, perception, productivity, psychology, physical and biological factors, health and reliability seem to stand out in functional processes how an individual perceives its environment. In maximizing the level of implementation of human productivity, technology and human relationships gain importantance. Behavioral and cognitive characteristics of people, you need to consider capabilities. Rules and approaches that will allow to make relationship between human behavioral, cognitive features, capabilities and technology the most favorable should be taken into account. Because, the usability vehicle is the software. For example, the model in the Table 1 shows the relevance between the characteristics brain and life skills of human and timescale (International Ergonomics Assosiation).

Table 1. The correlation between time and the quality of the human brain and life skills

| Qualification of the Human | Mean | Interval |
|---|---|---|
| The perceive time of eye | 230 ms | 70-700 ms |
| Half time of the image storage | 200 ms | 90-1000 ms |
| Turnaround time of discernment processor | 100 ms | 50-200 ms |
| The return period cognitive handler | 70 ms | 25-170 ms |
| Return time of move handler | 70ms | 30-100 ms |
| Efficient memory storage capacitance | 7 nesne | 5-9 nesne |

Human is one of the basic elements of software interface so how to be effective of human factor in usability process of cognitive processing characteristics is easily seen from the data in Table 1 [23,24]. When comes to the HCI issues; HCI is a field that examines the interaction of users and technology interfaces. At the same time, HCI aims to design, developing and evaluating to increase user efficiency and satisfaction of computer-based interactive systems [25-27]. In research on the communication process between the people and the system, technics, design, operating systems, programming languages, etc. are carried out in the system side, and communication theory, social sciences, linguistics, cognitive psychology, etc. are in the human side [28].

The main goal is, in context of particular using, to activate the interaction between users and product interfaces when performing specific tasks and to develop software and hardware that are compatible with users' needs, useful and user-friendly [25,26]. In human and computer interaction, the user and the system (computer) intercommunicate. In this communication, the user interacts in order to perform tasks in accordance with their goals and needs, while the system interacts to mediate between the user and tasks. Such interaction utilizes the user-centered design approach for the users to use the system more effectively and to learn more easily [25].

### 2.2.1. User-Centered Design

User-Centered Design (UCD) approach, is an interactive system development approach focused on increasing system interfaces' usability in line with user needs and expectations. In UCD, unlike in system-focused designs, the users are directly involved in the designing process. The aim is to enhance system's usability, usefulness and accesibility with the information acquired from the users. Norman and Draper (1986) emphasize that systems are designed for users, therefore users' needs must be at the forefront when designing interfaces. The methods used in the design of software applications are mostly concerned with technical requirements. However, user requirements have the same importance as the technical and functional requirements for a software. UCD aims to reflect the user's perspective on usable system designs [25, 29, 31].

According to International Organization for Standardization's (ISO) standard number 13407 named UCD processes for Interactive Systems, UCD is a process which is executed as "planning, development, measurement and implementation" and its design cycle is shown in Figure 2.

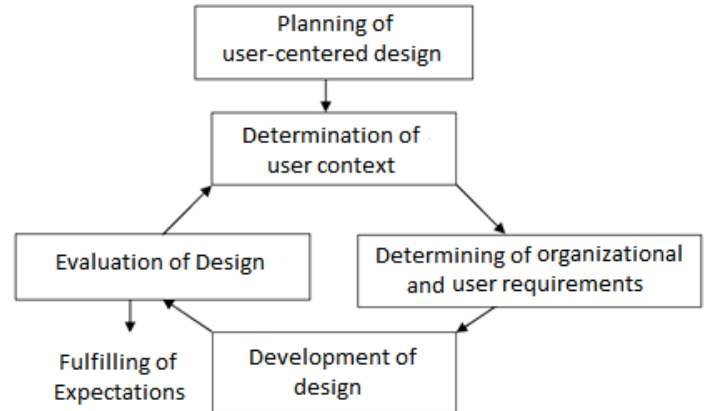

Figure 2. ISO 13407 User-Centered Design Cycle [25, 32]

ISO 13407 states that; UCD process, which is an interdisciplinary activity unifying human factors and information of ergonomics in order to improve effectiveness and efficiency of the system interface, requires a repetitive design process [25, 32].

Aside the use of interdisciplinary methods, adding a new layer into software technology emerges as a necessity. For this reason, the execution process in a complicated field such as bringing together the cognitive and behavioral features of human and the tools and special rules of software continues to advance as a sub-branch in software engineering by the multi-purpose name of "Usability Engineering". Viewed in the scope of M. Karat's determinations, the context of "usability" as a nodal point of factor that comes out with change is seen to be a required concessive method [23].

### 2.2.2. Usability

It is possible to find lots of definitions about usability in the field [29, 33, 35]. By ISO's definition, usability is users' state of effectiveness, efficiency and satisfaction about a developed system to achieve certain goals in a certain environment [33, 36, 38]. An acknowledged scientist in this field, Nielsen, defines usability as a combination of factors (easy to learn, efficient to use, rememberability, low error rate and satisfaction of use) which affect user's interaction with a product or a system [39]. In addition, Jones defines usability as the total effort required to learn, execute and use a software/hardware [33, 40, 42]. Brinck, Gergle and Wood define usability through various elements such as functionality, usage efficiency, ease of usage and learning, error toleration and favorability by the users [33, 43]. Thus, in a general sense, the purpose of usability is to design products appropriately for user expectations and needs and to increase the product's effectivity, efficiency and satisfaction degree [25, 33, 36].

### 3. Results and Suggestions

In this study, HCI's importance in SDP is examined. Significant results are obtained in the study which aims to reveal HCI's importance in SDP and thus to create an awareness on the subject. According to the results, especially in the software life-cycle, it is seen that the rules of interaction are implemented in the software life-cycle usually on the late stages of the software developing process, whereas these rules are to be prepared before the software development. As a result, it is clearly revealed that when the design methods used within the context of HCI are integrated into SDP during the software life-cycle, the developed software projects become more successful, higher quality and more user-friendly. Also, it must not be forgotten that user is the focal point when developing a software.

Therefore, user's experiences, abilities, participation, feed-back, satisfaction, behavioral and cognitive features and limitations must be considered. In this way, when generating usage efficiency of the software, the contributing causes of the software being taken into consideration in this process will become an important element in the quality of the developed software.